# Heralded generation of vectorially structured photons with high purity

Hai-Jun Wu,[1] Bing-Shi Yu,[1] Zhi-Han Zhu,[1,*] Carmelo Rosales-Guzmán,[1,2] Zhi-Yuan Zhou,[1,3] Dong-Sheng Ding,[1,3] Wei Gao[1] and Bao-Sen Shi[1,3,†]

[1] *Wang Da-Heng Collaborative Innovation Center, Heilongjiang Provincial Key Laboratory of Quantum Control, Harbin University of Science and Technology, Harbin 150080, China*

[2] *Centro de Investigaciones en Óptica, A.C., Loma del Bosque 115, Colonia Lomas del campestre, 37150 León, Gto., Mexico*

[3] *CAS Key Laboratory of Quantum Information, University of Science and Technology of China, Hefei, 230026, China*

Engineering vector spatial modes of photons is an important approach for manipulating high-dimension photonic states in various quantum optical experiments. In this work, we demonstrate generation of heralded single photons with well-defined vector spatial modes by using a self-locking polarizing interferometer comprising a spatial light modulator. Specifically, it is shown that, by carefully tailoring and compensating spatial and temporal amplitudes of manipulated photons, one can exactly convert ultrafast single photons into desired spin-orbit states with extremely high purity. This compact and robust device provides a versatile way for not only generation, but also manipulation and characterization of arbitrary photonic spin-orbit states.

Vectorially structured photons with well-defined vector spatial modes are paraxial light fields whose polarization and transverse modes are non-separable with each other, also known as spin-orbit coupled (SOC) states [1-4]. Exploiting photonic SOC states can benefit many quantum technologies based on photonic platforms, such as, high-dimension quantum information [5-9]. To explore this area, one requires on-demand control of both spin and orbital degrees of freedom (DoFs) of photons. Namely, having an ability of polarization-dependent spatial mode modulation. The most convenient way for this task is the use of optical geometric phase elements, such as q-plates [10,11]. To date, however, all these elements can only provide phase-only modulation lacking the ability to modulate the spatial complex amplitude of light fields. Therefore, these devices can only work to generate approximate cylindrical vector (CV) modes [12,13], characterized by a series of propagation-variant radial rings, whose spatial parts are more precisely described as radial-mode degenerated orbital angular momentum (OAM) modes [14]. Noteworthy, to simultaneously shape and control the phase and amplitude of vectorially structured photons, the only feasible way at present is to use spatial light modulators (SLM) combined with various polarizing interferometer schemes.

For example, M. Christian et al proposed a self-locking interferometer scheme, based on the combination of a Wollaston prism and an SLM, capable to generate arbitrary vector spatial modes using complex amplitude modulation implemented on the SLM [15]. Yet, more than 75% photons will be lost due to double passed a beam splitter, and moreover, the calcite made Wollaston prism reduced the beam quality. Recently, Ref. 16 reported that using two polarizing Sagnac loops as tunable beam displacers and a phase-only SLM realized arbitrary vector mode generation with a 47% conversion efficiency. This setup, however, is not compact enough as a SOC module for quantum optical experiments, and they did not demonstrate generation of propagation invariant modes via this device. More recently, we proposed a compact DMD based setup for dynamical generation of arbitrary vector modes with a 30 kHz frame rate [17]. Given that DMDs can only realize binary holographic grating, the photon utilization of this device is too low for quantum experiments.

In this work, we report a compact and robust device for generation, manipulation, and characterization of vectorially structured photons. To demonstrate its performance, an ultrafast single-photon signal with $TEM_{00}$ mode was efficiently converted into a series of vector spatial modes with high accuracy. Particularly, we show that the purity of prepared SOC states can get extremely close to the theoretical predictions, after compensating the errors in the spatial and temporal amplitudes of converted photons.

Vector spatial modes are theoretical eigen-solutions of the vector paraxial wave equation that corresponds to propagation invariant beams with spatially variant polarization. These vector solutions can also be expressed as a non-separable superposition of orthogonal scalar spatial modes $\psi_\pm$ and associated polarizations $\hat{e}_\pm$, which in cylindrical coordinates are given by $E(r,\varphi,z) = \alpha\psi_+(r,\varphi,z)\hat{e}_+ + \beta\psi_-(r,\varphi,z)\hat{e}_-$, where $\alpha$ and $\beta$ are complex probability amplitudes. Importantly, it is possible to represent any spatial mode as a superposition of OAM carrying modes, e.g., Laguerre-Gauss (LG) modes. That is, vector spatial modes can also be generally regarded as photonic SOC states, and thus, can be simply denoted using the Dirac notation as $\alpha|\psi_+,\hat{e}_+\rangle+\beta|\psi_-,\hat{e}_-\rangle$. To experimentally generate and control this photonic state, any apparatus must be capable to simultaneously tailor the spatial complex amplitudes of the two spin-dependent spatial modes.

* zhuzhihan@hrbust.edu.cn
† drshi@ustc.edu.cn

Figure 1 shows a schematic representation of our proposed device, which is a compact self-locking Mach-Zehnder interferometer. The key components comprise a pair of polarizing beam displacing (PBD) prisms made by fused silica. Compared to calcite beam displacer (as the one used in [15]), the PBD features higher transmission, low beam distortion, and the option for a flexible separation-distance. By contrast, the length of a calcite beam displacer with a 5 mm separation distance will exceed 5 cm, leading to a serious beam distortion and high cost. To experimentally demonstrate the feasibility of our device for tailoring ultrafast single photons, we used a 1.5 mm type-II PPKTP crystal pumped by a SHG of 795nm laser (Toptica TA Pro) to generate the heralded single-photon signal. The photon source has a 5 MHz brightness (pairs/second) with a heralding ratio 90% (detector excluded). The measured HOM dip of the photon pair is shown in the upper-left inset in Fig. 1, where a 340 fs FWHM indicates the coherent time of photons is approximately 240 fs. To generate a desired SOC state $\alpha|\psi_H,\hat{e}_H\rangle+\beta|\psi_V,\hat{e}_V\rangle$, the single-photon signal, which enters from the left side of the setup, was converted into a polarization-path non-separable state, i.e., $\alpha|\hat{e}_H,01\rangle+\beta|\hat{e}_V,10\rangle$, by a half-wave plate (HWP-1) and a PBD-1. After that, each single photon was simultaneously sent to two different sections, independently controlled, of a high reflectance (99%) SLM (Holoeye PLUTO-2-080). In addition, the probability amplitude of photons with V-polarization, i.e., $\beta$, twice passed through the HWP-2 fixed at 45°, this is required because SLMs can only modulate H-polarized light. Hence, by independently tailoring the spatial mode within two polarization subspaces of the photons, they were further converted into a three DoFs non-separable state, i.e., $\alpha|\psi_H,\hat{e}_H,01\rangle+\beta|\psi_V,\hat{e}_V,10\rangle$. Finally, HWP-3 and PBD-2 recombined the single photons travelling along the two paths [18], generating in this way the desired SOC state. Additionally, a single-photon sensitive camera (EMICCD, PI-MAX4) combined with polarizers were used to perform spatial Stokes tomography on the generated modes [19], as shown in the upper-right inset of Fig. 1.

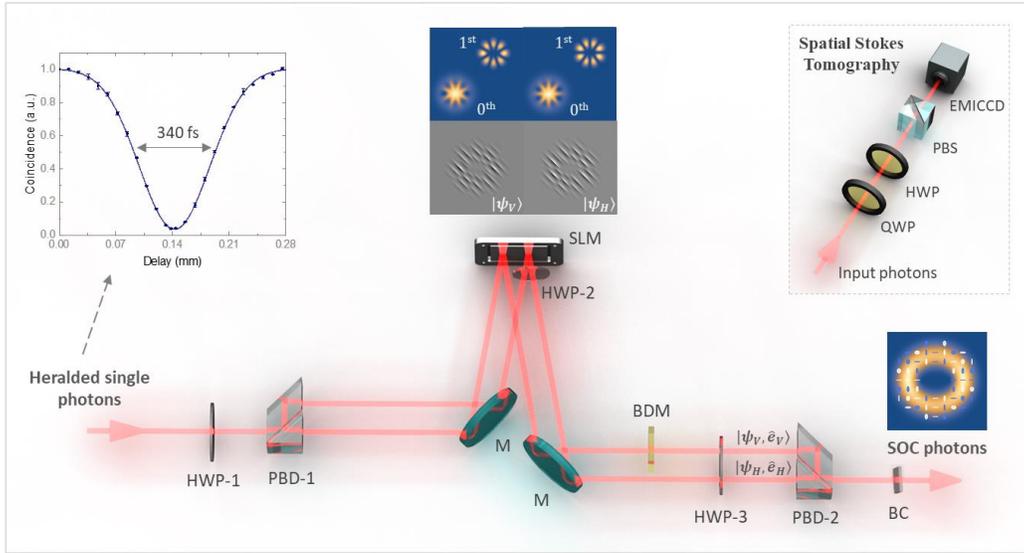

FIG. 1. Schematic representation of the experimental setup, where the key components include the polarizing beam displacing prism (PBD), a spatial light modulator (SLM), half-wave plates (HWP), a beam displacement module (BDM), mirrors (M), and a birefringence crystal (BC). The upper-left inset shows the HOM dip of two-photon pairs, the upper-right one shows the setup for the spatial Stokes tomography.

The purity of the generated SOC states lies with the consistency of complex amplitude, in both spatial and temporal DoFs, between the prepared and the desired state. Experimentally, realization of this consistency requires high accuracies in orthogonal polarization control, spatial modes generation, and the installation and adjustment of the interferometer. For the polarization control, the fused silica PBD can provide a high enough extinction ratio (>1000:1). Therefore, the other two factors are crucial for the generation of pure states.

The principle of complex amplitude modulation via a phase-only SLM consist on designing a special phase hologram with blazed grating whose spatial phase-delay and grating depth (efficiency) control the wavefront and the intensity of diffracted light, respectively. The corresponding desired spatial mode is generated along the 1st order diffraction [20]. Note that the input photons are not ideal plane wave, usually a Gauss mode. Therefore, to generate a well-defined spatial mode, we need to make a correction in the hologram according to the intensity profile of input photons.

Figure 2(a) shows the observed beam profile of input photons before (OI) and after (FI) using Fourier noise reduction, as well as the case of maximum photon utilization for generating $LG_{01}$ mode (TI). For comparison, Fig. 2(b) shows the holograms before and after the correction and associated diffraction results. We see that, by using the corrected hologram, an exact $LG_{01}$ mode appears in the 1$^{st}$ order diffraction. Besides, Fig. 2(c) provides generation efficiency for various modes, where $LG$ and $IG$ denote Laguerre- and Ince-Gauss modes [21-25], respectively. We see that, for all cases, the measured efficiencies approach the theoretical limitation.

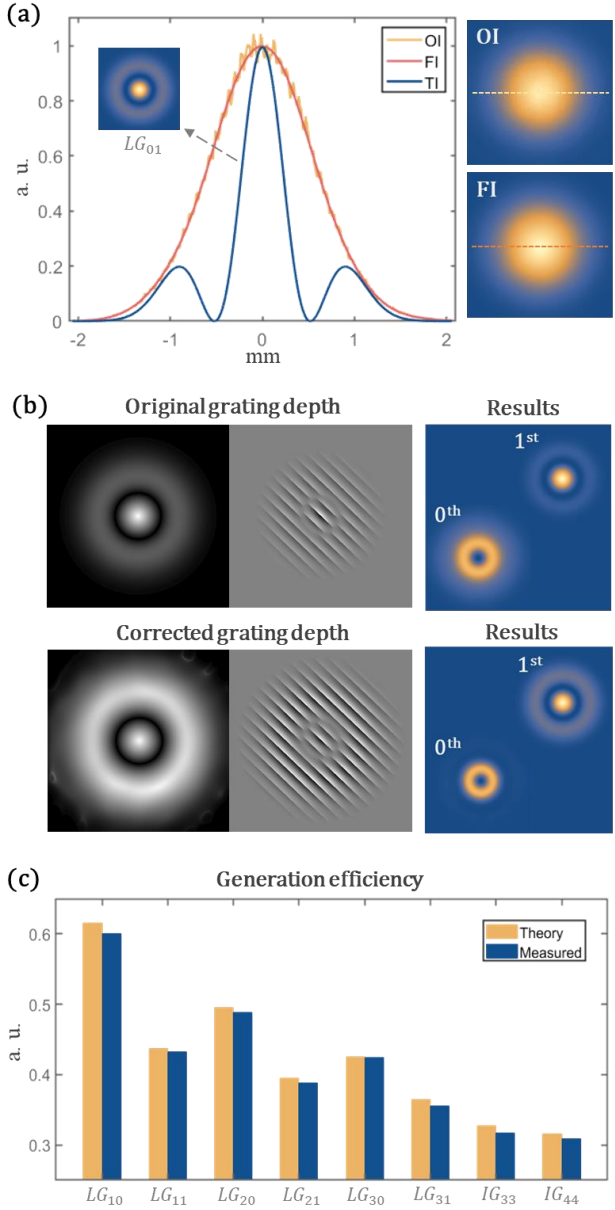

FIG. 2. Results of spatial mode generation, (a) characterization of illuminating photons; (b) correction of the hologram for spatial mode generation; (c) generation efficiencies for different spatial modes about the 1$^{st}$ order diffraction.

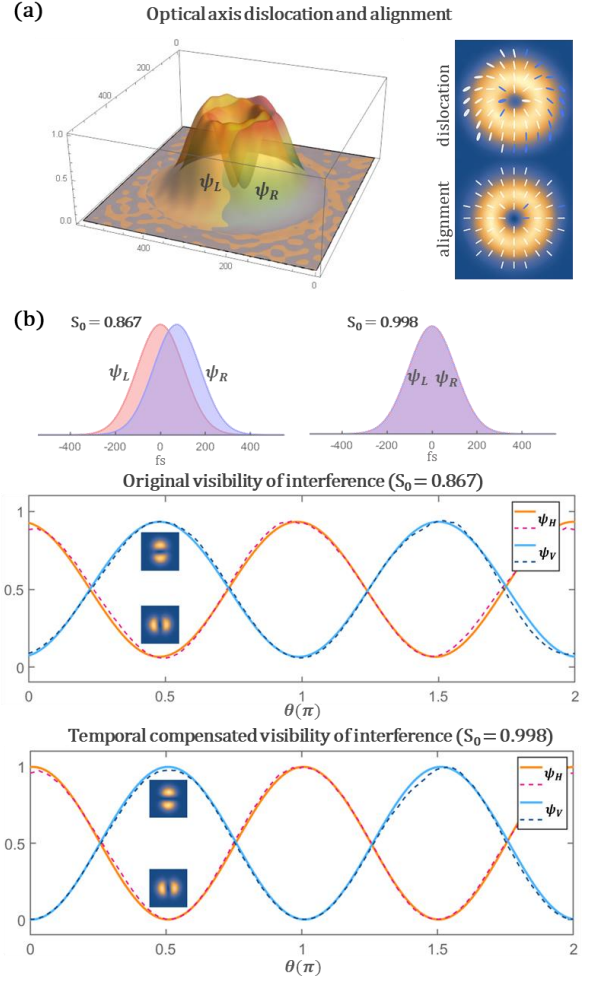

FIG. 3. Results of the compensation of transverse (a) and temporal (b) dislocations.

In the interferometer adjustment, it is important to note that there are errors in the separation distance and parallelism of the PBD. Namely, using the prim pairs cannot direct construct a perfect beam separation and combination loop, leading to a dislocation between the optical axis of $\psi_\pm$. To compensate this dislocation, two techniques were adopted: (i) a 2 mm tweaker plate was used as a beam displacement module (BDM) to compensate the error of PBD in separation distance; and (ii) fine control the grating period of the arm inserted BDM so that the error of PBD in parallelism can be well compensated. For comparison, Fig. 3(a) shows the observed radial polarization mode, i.e., $\sqrt{1/2}(|\psi_L, \hat{e}_R\rangle + |\psi_L, \hat{e}_R\rangle)$, before and after the compensation of spatial amplitude.

In addition, because the coherent length of input photons is only about 72 μm, the temporal dislocation between $\psi_\pm$ accumulated in the interferometer should be considered and compensated. The optical path difference (OPD) between the two arms induced by transmission components is small (μm level), because the thickness of HWP-2 (double passed) and BDM are 1 mm and 2 mm, respectively. The OPD between two arms is mainly originated from the extraordinary reflection angle induced by the blazed grating. In the experiment, the grating period is 200 μm, the incident angle is within 5 degree, and the beam distance is 5 mm. By calculation, we known the OPD induced by the blazed grating was approximate 20 μm. This temporal dislocation can be easily compensated by passing a birefringence crystal, such as a 0.2 – 0.3 mm BBO or KTP crystal.

In order to accurately know the OPD, we measured the azimuthal interference visibility of the transverse pattern of a prepared radial-polarization CV mode in $\hat{e}_H$- and $\hat{e}_V$-postselection, i.e., $\psi_{H/V}(\theta) = \sqrt{1/2}(|LG_{+10}\rangle \pm e^{i\theta}|LG_{-10}\rangle)$, as shown in Fig. 3(b). By calculating the average Stokes parameter $S_0$ over the transverse plane, i.e., $\bar{S}_0(x,y)$, we known that the visibility was 0.867 corresponding to a 32 μm OPD. Then, we inserted a 300 μm BBO crystal at the output port to compensate this temporal dislocation. As shown in Fig. 3(b), a near perfect pure state with an extremely high 0.998 visibility was achieved after the temporal dislocation compensation.

After finishing adjustment of the interferometer, we choose two groups of vectorially structured photons generated by the device to show its performance, as shown in Fig. 4. Specifically, the high-order Poincaré sphere in the first group is spanned by $|LG_{+21},\hat{e}_L\rangle$ and $|LG_{-21},\hat{e}_L\rangle$, while in the other one $\psi_{L/R} = \sqrt{1/2}(|IG_{44}^0\rangle \pm i|IG_{44}^e\rangle)$ with an ellipticity $\varepsilon = 1$. We see that the vector profiles of prepared SOC states excellently coincide with the theoretical desired states. Particularly, the modes in the second group are vector Ince-Gauss modes that were proposed in our recent work [25]. These modes are propagation invariant and thus can be regard as eigen solutions of the vector paraxial wave equation in the corresponding elliptical cylindrical coordinates.

We have demonstrated the superiority of our device for generation of arbitrary high-purity SOC states of single photons. Noteworthy, this device can be also used for shaping entangled photons, for example, converting a polarization entanglement into a hyperentanglement and vice vera. Besides, in classical domain, this device has also advantage in efficiency and accuracy for shaping vector beam or pulses compared to previous works [15,16]. In summary, we proposed a SLM-based self-locking polarization interferometer for shaping vectorially structured photons. By theoretical analysis and experimental observation, we show that this compact and robust device can efficiently convert ultrafast photons into desired SOC states with extremely high purity. This programmable interferometer can also work as a versatile photonic SOC modulator for manipulation and characterization of photonic SOC states. During the preparation of this paper, we found a preprint paper reporting the similar topic, where a more compact prism-based interferometer was adopted [26]. However, due to the absence of SLM, this prism interferometer can only work for controlling CV modes. Namely, this prism-based interferometer can be regarded as an integrated version of the Sagnac loop used in our recent experiments [27-29].

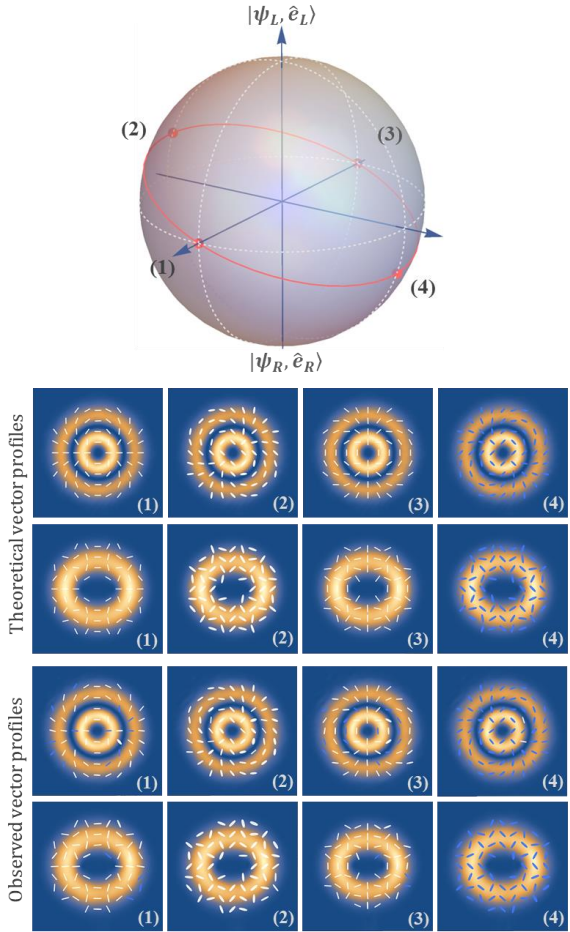

FIG. 4. Comparison of generated and desired SOC states on high-order Poincaré sphere, where the vector profiles were obtained by spatial Stokes tomography with 10s on-chip accumulation and Fourier noise reduction.

### Funding

This work was funded by the National Natural Science Foundation of China (Grant Nos. 11934013, 62075050, and 61975047).